\documentclass[a4paper,11pt]{article}
\usepackage{amssymb}
\usepackage{pos}
\usepackage{longtable}
\usepackage{booktabs,colortbl}
\usepackage{multirow}
\usepackage{amsmath}
\usepackage{bbm}
\usepackage[utf8]{inputenc}
\usepackage{pdflscape}
\usepackage{xspace}
\usepackage[caption=false]{subfig}
\usepackage{nicefrac}
\usepackage{graphicx}
\usepackage{mathtools}
\usepackage{nccfoots}
\usepackage{tikz}
\usetikzlibrary{calc,positioning}
\usepackage{bm}
\newlength{\bibitemsep}
\newlength{\bibparskip}
\let\oldthebibliography\thebibliography
\renewcommand\thebibliography[1]{%
  \oldthebibliography{#1}%
  \setlength{\parskip}{\bibitemsep}%
  \setlength{\itemsep}{\bibparskip}%
}

\usepackage[all,cmtip]{xy}
\newlength{\xywd}
\newcommand{\xyrightarrow}[2][]{%
  \sbox{0}{$\scriptstyle#1$}%
  \xywd=\wd0
  \sbox{0}{$\scriptstyle#2$}%
  \ifdim\wd0>\xywd \xywd=\wd0 \fi
  \xymatrix@C\dimexpr\xywd+1em\relax{{}\ar[r]^{#2}_{#1}&{}}%
}

\DeclareMathOperator{\re}{Re}
\DeclareMathOperator{\im}{Im}

\newcommand{\rep}[1]{\ensuremath\boldsymbol{#1}}
\newcommand{\crep}[1]{\ensuremath\bar{\boldsymbol{#1}}}
\newcommand{\Z}[1]{\ensuremath{\mathbbm{Z}_{#1}}} % z_N ->\Z{N}

\newcommand{\SL}[1]{\ensuremath{\mathrm{SL}(#1)}}

\newcommand{\I}{\mathrm{i}}

\newcommand{\CP}{\ensuremath{\mathcal{CP}}\xspace}

\newcommand{\vev}[1]{\ensuremath{\langle{#1}\rangle}}

\usepackage{xcolor}
\definecolor{darkgreen}{HTML}{109930}
\definecolor{pink}{rgb}{0.858, 0.188, 0.478}

\title{Modular Flavor Symmetries and CP from the top down}
\author*[a]{Andreas Trautner}
\affiliation[a]{Max-Planck-Institut f{\"u}r Kernphysik \\ Saupfercheckweg 1, 69117 Heidelberg, Germany}

\emailAdd{trautner@mpi-hd.mpg.de}
\abstract{%
The framework of compactified heterotic string theory offers consistent ultraviolet (UV) completions of the Standard Model (SM) of particle physics. 
In this approach, the existence of flavor symmetries beyond the SM is imperative and the flavor symmetries can be derived from the top down. 
Such a derivation uncovers a unified origin of traditional discrete flavor symmetries, discrete modular flavor symmetries, discrete R symmetries of supersymmetry, as well as charge-parity (CP) symmetry -- altogether dubbed the eclectic flavor symmetry.
I will illustrate how the eclectic flavor symmetry is unambiguously computed from the top-down construction, discuss the different arising sources of spontaneous flavor symmetry breaking, 
and expose possible lessons for bottom-up flavor model building. 
Finally, I will focus on one explicit example model that provides a successful fit to all available experimental data while giving rise to concrete predictions for so-far undetermined parameters.}

\FullConference{%
  8th Symposium on Prospects in the Physics of Discrete Symmetries (DISCRETE 2022)\\
  7-11 November, 2022\\
  Baden-Baden, Germany
}

\begin{document}
\maketitle

\section{Introduction}
Embracing grand unification, a solution to the electroweak hierarchy problem as well as a consistent quantum theory of gravity, 
there is strong motivation from the bottom-up to consider string theory as UV completion of the SM. 
At the same time, it is crucial to ensure that the SM can indeed be consistently incorporated in a concrete realization of string theory and expose
the constraints and predictions that may arise from such a derivation.
A particularly advanced setup is provided by orbifold compactifications of the heterotic string \cite{Ibanez:1986tp,Ibanez:1987sn,Gross:1984dd,Gross:1985fr,Dixon:1985jw,Dixon:1986jc} 
which cannot only consistently host the (supersymmetric) SM~\cite{Buchmuller:2005jr} but may also shed light on one of its
most pressing puzzles by automatically including family repetition and flavor symmetries~\cite{Kobayashi:2006wq,Olguin-Trejo:2018wpw}.
Since a ``theory of everything'' has to be, in particular, a theory of flavor, there is a strong motivation
to understand the flavor puzzle of the SM from such a top-down perspective.

In this talk (see~\cite{Trautner:2022biw} for an earlier version), we present new progress achieved in consistently deriving 
the complete flavor symmetry from concrete string theory models, including the role of the different sources 
that can contribute to the flavor symmetry breaking in the infrared (IR).
As a proof of principle, we present the first consistent string theory derived model that gives rise to potentially realistic low-energy 
flavor phenomenology~\cite{Baur:2022hma}. The example is a heterotic string theory compactified on a $\mathbbm{T}^2/\Z3$ manifold. Our results show that
this model provides a successful fit to all available experimental data while giving rise to concrete predictions for so-far undetermined parameters.
Corrections from the K\"ahler potential turn out to be instrumental in obtaining a successful simultaneous fit to quark and lepton data.
While in our effective description there are still more parameters than observables, this gives a proof of principle of the existence of consistent global explanations of flavor in the quark and lepton sector from a top-down perspective.
In the end we will also point out possible lessons for bottom-up flavor model building and important open problems. 

\section{Types of discrete flavor symmetries and the eclectic symmetry}
The action of the 4D effective $\mathcal{N}=1$ SUSY theory
can schematically be written as (here, $K$: K\"ahler potential, $W$: Superpotential, $x$: spacetime, $\theta$: superspace, $\Phi:$ superfields, $T$: modulus)
\begin{equation}
\mathcal{S}=\int d^4x\,d^2\theta\,d^2\bar{\theta}\,K(T,\bar{T},\Phi,\bar{\Phi})+ \int d^4x\,d^2\theta\,W(T,\Phi)+ \int d^4x\,d^2\bar{\theta}\,\bar{W}(\bar{T},\bar{\Phi})\;.
\end{equation}
There are four categories of possible symmetries that differ by their effect on fields and coordinates:
\begin{itemize}
 \item ``Traditional'' flavor symmetries ``$G_\mathrm{traditional}$'', see e.g.~\cite{Ishimori:2010au}: $\Phi\mapsto \rho(\mathsf{g})\Phi$\,,\quad$\mathsf{g}\in G_\mathrm{traditional}$.
  \item Modular flavor symmetries ``$G_\mathrm{modular}$'' \cite{Feruglio:2017spp}: (partly cancel between $K$ and $W$)   
 \begin{equation}
\gamma:=\begin{pmatrix}a&b\\c&d\end{pmatrix}\in\mathrm{SL}(2,\Z{})\;,\quad\Phi\xmapsto{\gamma}(c\,T+d)^n \rho(\gamma)\Phi\;,\quad  T\xmapsto{\gamma}\frac{a\,T+b}{c\,T+d}\;.  
 \end{equation}
 In this case couplings are promoted to modular forms: $Y=Y(T)$, $Y(\gamma T)=\left(c\,T+d\right)^{k_Y}\rho_Y(\gamma)\,Y(T)$.
 \item R flavor symmetries ``$G_R$'' that differ for fields and their superpartners~\cite{Chen:2013dpa}~(cancel between $W$ and $d^2\theta$).
 \item General symmetries of the ``\CP'' type~\cite{Baur:2019kwi,Novichkov:2019sqv}: (partly cancel between $K$ and $W$ and $d^4x$)
 \begin{equation}
  \det\left[\,\bar{\gamma}\in\mathrm{GL}(2,\Z{})\,\right]=-1\;,\quad\Phi~\xmapsto{\bar{\gamma}}~(c\bar{T}+d)^n \rho(\bar{\gamma}) \bar{\Phi}\;,\quad  T~\xmapsto{\bar{\gamma}}~\frac{a\bar{T}+b}{c\bar{T}+d}\;.
 \end{equation}
\end{itemize}
All of these symmetries are individually known from bottom-up model building, see~\cite{Feruglio:2019ybq}. In explicit top-down constructions
we find that \textit{all of these arise at the same time} in a non-trivially unified fashion~\cite{Baur:2019iai,Baur:2019kwi,Nilles:2020nnc,Nilles:2020kgo,Nilles:2020tdp,Nilles:2020gvu,Ohki:2020bpo,Nilles:2021ouu}, that we call the ``eclectic'' flavor symmetry~\cite{Nilles:2020nnc}
\begin{equation}\label{eq:eclectic}
\fcolorbox{red}{yellow}{\Large ~$G_\mathrm{\mathbf{eclectic}} ~=~ G_\mathrm{traditional} ~\cup~ G_\mathrm{modular} ~\cup~ G_\mathrm{R} ~\cup~ \CP\,.$\vspace{0.2cm}~}
\end{equation}

\section{Origin of the eclectic flavor symmetry in heterotic orbifolds}
A new insight is that in the Narain lattice formulation of compactified heterotic string theory~\cite{Narain:1985jj,Narain:1986am,GrootNibbelink:2017usl}
the complete unified eclectic flavor symmetry can unambiguously derived from the outer automorphisms~\cite{Trautner:2016ezn} of the Narain lattice space group~\cite{Baur:2019kwi, Baur:2019iai}.
These outer automorphisms contain modular transformations, including the well-known T-duality transformation and the so called mirror symmetry (permutation of different moduli) of string theory, but also symmetries of the \CP-type as well as traditional flavor symmetries and, therefore, naturally yield the unification shown in Eq.~\eqref{eq:eclectic}.
The eclectic transformations also automatically contain the previously manually derived so-called ``space-group selection rules'' \cite{Hamidi:1986vh,Dixon:1986qv,Ramos-Sanchez:2018edc} 
and non-Abelian ``traditional'' flavor symmetries~\cite{Kobayashi:2006wq}.

\section[The eclectic flavor symmetry of T2/Z3]{\boldmath The eclectic flavor symmetry of $\mathbbm T^2/\Z3$ \unboldmath}
Let us now focus on a specific example model~\cite{Baur:2021bly} in which the six extra dimensions of ten-dimensional heterotic string theory are compactified in such
a way that two of them obey the $\mathbbm T^2/\Z3$ orbifold geometry.  The discussion of this $D=2$ subspace involve a K\"ahler and complex structure modulus
$T$ and $U$, respectively, with the latter being fixed to $\langle U\rangle=\exp(\nicefrac{2\pi\I}3)=:\omega$ by the orbifold action.
The outer automorphisms of the corresponding Narain space group yield the full eclectic group of this setting, which is of order $3888$ and given by\footnote{%
Finite groups are denoted by $\mathrm{SG}\left[\cdot,\cdot\right]$ where the first number is the order of the group and the second their GAP SmallGroup ID~\cite{GAP4url}.
}~\cite{Nilles:2020tdp,Nilles:2020gvu}
\begin{equation}
  G_\mathrm{eclectic} ~=~ \Omega(2)\rtimes\Z2^\CP\,,\qquad\text{with}\quad \Omega(2)\cong \mathrm{SG}[1944,3448]\,.
\end{equation}
More specifically, $G_\mathrm{eclectic}$ contains
\begin{itemize}
\item a $\Delta(54)$ traditional flavor symmetry,
\item the $\mathrm{SL}(2,\Z{})_T$ modular symmetry of the $T$ modulus, which acts as a $\Gamma'_{3}\cong T'$ finite 
      modular symmetry on matter fields and their couplings,
\item a $\Z9^R$ discrete R symmetry as remnant of $\mathrm{SL}(2,\Z{})_U$, and 
\item a $\Z2^\CP$ \CP-like transformation.
\end{itemize}
These symmetries and their interplay are shown in table~\ref{tab:Z3FlavorGroups}. 
Twisted strings localized at the three fixed points of the $\mathbbm{T}^2/\Z3$ orbifold
form three generations of massless matter fields in the effective IR theory with 
transformations under the various symmetries summarized in table~\ref{tab:Representations}.
Explicit representation matrices of the group generators are shown in the slides of the talk and in the papers~\cite{Baur:2021bly,Baur:2022hma}.
Examples for complete string theory realizations are known, see~\cite{Carballo-Perez:2016ooy,Ramos-Sanchez:2017lmj} and \cite{Baur:2021bly,Baur:2022hma}, 
and we show the derived charge assignment of the SM-like states in one particular example in table~\ref{tab:Z3xZ3configurations}.
\begin{table}[t!]
\center
\resizebox{\textwidth}{!}{
\begin{tabular}{|c|c||c|c|c|c|c|c|}
\hline
\multicolumn{2}{|c||}{nature}        & outer automorphism       & \multicolumn{5}{c|}{\multirow{2}{*}{flavor groups}} \\
\multicolumn{2}{|c||}{of symmetry}   & of Narain space group    & \multicolumn{5}{c|}{}\\
\hline
\hline
\parbox[t]{3mm}{\multirow{6}{*}{\rotatebox[origin=c]{90}{eclectic}}} &\multirow{2}{*}{modular}            & rotation $\mathrm{S}~\in~\SL{2,\Z{}}_T$ & $\Z{4}$      & \multicolumn{3}{c|}{\multirow{2}{*}{$T'$}} &\multirow{6}{*}{$\Omega(2)$}\\
&                                    & rotation $\mathrm{T}~\in~\SL{2,\Z{}}_T$ & $\Z{3}$      & \multicolumn{3}{c|}{}                      & \\
\cline{2-7}
&                                    & translation $\mathrm{A}$                & $\Z{3}$      & \multirow{2}{*}{$\Delta(27)$} & \multirow{3}{*}{$\Delta(54)$} & \multirow{4}{*}{$\Delta'(54,2,1)$} & \\
& traditional                        & translation $\mathrm{B}$                & $\Z{3}$      &                               & & & \\
\cline{3-5}
& flavor                             & rotation $\mathrm{C}=\mathrm{S}^2\in\SL{2,\Z{}}_T$      & \multicolumn{2}{c|}{$\Z{2}^R$} & & & \\
\cline{3-6}
&                                    & rotation $\mathrm{R}\in\SL{2,\Z{}}_U$   & \multicolumn{3}{c|}{$\Z{9}^R$}   & & \\
\hline
\end{tabular}
}
\caption{\label{tab:Z3FlavorGroups}
Eclectic flavor group $\Omega(2)$ for six-dimensional orbifolds that contain a 
$\mathbbm T^2/\Z{3}$ orbifold sector~\cite{Nilles:2020tdp}. 
}
\end{table}
\begin{table}[h]
\center
\resizebox{\textwidth}{!}{
\begin{tabular}{|c||c||c|c|c|c||c|c|c|c||c|}
\hline
\multirow{3}{*}{sector} &\!\!matter\!\!& \multicolumn{9}{c|}{eclectic flavor group $\Omega(2)$}\\
                        &fields        & \multicolumn{4}{c||}{modular $T'$ subgroup} & \multicolumn{4}{c||}{traditional $\Delta(54)$ subgroup} & $\Z{9}^R$ \\
                        &$\Phi_n$      & \!\!irrep $\rep{s}$\!\! & $\rho_{\rep{s}}(\mathrm{S})$ & $\rho_{\rep{s}}(\mathrm{T})$ & $n$ & \!\!irrep $\rep{r}$\!\! & $\rho_{\rep{r}}(\mathrm{A})$ & $\rho_{\rep{r}}(\mathrm{B})$ & $\rho_{\rep{r}}(\mathrm{C})$ & $R$\\
\hline
\hline
bulk      & $\Phi_{\text{\tiny 0}}$   & $\rep1$             & $1$                   & $1$                   & $0$               & $\rep1$   & $1$               & $1$                   & $+1$ & $0$         \\
          & $\Phi_{\text{\tiny $-1$}}$& $\rep1$             & $1$                   & $1$                   & $-1$              & $\rep1'$  & $1$               & $1$                   & $-1$ & $3$         \\
\hline
$\theta$  & $\Phi_{\nicefrac{-2}{3}}$ & $\rep2'\oplus\rep1$ & $\rho(\mathrm{S})$    & $\rho(\mathrm{T})$    & $\nicefrac{-2}{3}$& $\rep3_2$ & $\rho(\mathrm{A})$& $\rho(\mathrm{B})$    & $+\rho(\mathrm{C})$ & $1$\\
          & $\Phi_{\nicefrac{-5}{3}}$ & $\rep2'\oplus\rep1$ & $\rho(\mathrm{S})$    & $\rho(\mathrm{T})$    & $\nicefrac{-5}{3}$& $\rep3_1$ & $\rho(\mathrm{A})$& $\rho(\mathrm{B})$    & $-\rho(\mathrm{C})$ & $-2$\\
\hline
$\theta^2$& $\Phi_{\nicefrac{-1}{3}}$ & $\rep2''\oplus\rep1$& $(\rho(\mathrm{S}))^*$& $(\rho(\mathrm{T}))^*$& $\nicefrac{-1}{3}$& $\crep3_1$& $\rho(\mathrm{A})$& $(\rho(\mathrm{B}))^*$& $-\rho(\mathrm{C})$ & $2$\\
          & $\Phi_{\nicefrac{+2}{3}}$ & $\rep2''\oplus\rep1$& $(\rho(\mathrm{S}))^*$& $(\rho(\mathrm{T}))^*$& $\nicefrac{+2}{3}$& $\crep3_2$& $\rho(\mathrm{A})$& $(\rho(\mathrm{B}))^*$& $+\rho(\mathrm{C})$ & $5$\\
\hline
\hline
super-    & \multirow{2}{*}{$W$} & \multirow{2}{*}{$\rep1$} & \multirow{2}{*}{$1$} & \multirow{2}{*}{$1$} & \multirow{2}{*}{$-1$} & \multirow{2}{*}{$\rep1'$} & \multirow{2}{*}{$1$} & \multirow{2}{*}{$1$} & \multirow{2}{*}{$-1$} & \multirow{2}{*}{$3$}\\
\!\!potential\!\! & & & & & & & & & & \\
\hline
\end{tabular}
}
\caption{\label{tab:Representations}
$T'$, $\Delta(54)$ and $\Z9^R$ representations of massless matter fields $\Phi_n$ with modular weights $n$ 
in semi-realistic heterotic orbifold compactifications with a $\mathbbm{T}^2/\Z3$ sector~\cite{Nilles:2020kgo}. 
}
\end{table}
\begin{table}[t]
	\centering
	%\resizebox{\textwidth}{!}{ 
		\begin{tabular}{cllllllllll}
			\toprule
			& $\ell$                   & $\bar e$              & $\bar\nu$             & $q$                   & $\bar u$
			& $\bar d$              & $H_u$                 & $H_d$                 &  $\varphi_\mathrm{f}$ & $\phi^0_\mathrm{f}$\\
			\midrule
			Model A & $\Phi_{\nicefrac{-2}3}$ & $\Phi_{\nicefrac{-2}3}$ & $\Phi_{\nicefrac{-2}3}$ & $\Phi_{\nicefrac{-2}3}$ & $\Phi_{\nicefrac{-2}3}$
			& $\Phi_{\nicefrac{-2}3}$ & $\Phi_{0}$  & $\Phi_{0}$  & $\Phi_{\nicefrac{-2}3}$ & $\Phi_{0}$\\
%			B & $\Phi_{\nicefrac{-1}3}$ & $\Phi_{\nicefrac{-2}3}$ & $\Phi_{\nicefrac{-2}3}$ & $\Phi_{\nicefrac{-2}3}$ & $\Phi_{\nicefrac{-2}3}$
%			& $\Phi_{\nicefrac{-1}3}$ & $\Phi_{-1}$ & $\Phi_{0}$  & $\Phi_{\nicefrac{-2}3,-1}$ \\
%			C & $\Phi_{\nicefrac{-2}3}$ & $\Phi_{\nicefrac{-1}3}$ & $\Phi_{\nicefrac{-1}3}$ & $\Phi_{\nicefrac{-1}3}$ & $\Phi_{\nicefrac{-1}3}$
%			& $\Phi_{\nicefrac{-2}3}$ & $\Phi_{-1}$ & $\Phi_{-1}$ & $\Phi_{\nicefrac{-1}3,-1}$\\
%			D & $\Phi_{\nicefrac{-1}3}$ & $\Phi_{\nicefrac{-1}3}$ & $\Phi_{\nicefrac{\pm2}3,0}$ & $\Phi_{\nicefrac{-1}3}$ & $\Phi_{\nicefrac{-1}3}$
%			& $\Phi_{\nicefrac{-1}3}$ & $\Phi_{0}$ & $\Phi_{-1,0}$ & $\Phi_{\nicefrac{\pm2}3,-1}$\\
%			E & $\Phi_{\nicefrac{-2}3,\nicefrac{-1}3}$ & $\Phi_{\nicefrac{-2}3,0}$ & $\Phi_{0,\nicefrac{-2}3,\nicefrac{-1}3,\nicefrac{-5}3}$ & $\Phi_{-1,\nicefrac{-2}3}$ & $\Phi_{\nicefrac{-2}3}$
%			& $\Phi_{0,\nicefrac{-2}3}$ & $\Phi_{0}$ & $\Phi_{0}$ & $\Phi_{\nicefrac{-2}3,\nicefrac{-1}3,\nicefrac{-5}3,-1}$\\
			\bottomrule
		\end{tabular}
	%}
	\caption{\label{tab:Z3xZ3configurations}
		Flavor symmetry representations of MSSM quark ($q,\bar u,\bar d$), lepton ($\ell,\bar e,\bar\nu$), Higgs and flavon fields ($\varphi$,$\phi$)
		in an example of a consistent string theory configuration with a $\mathbbm T^2/\Z3$ orbifold sector.
		Following the notation of table~\ref{tab:Representations}, representations are \textit{entirely} determined by stating the respective modular weight.
		The \textit{complete} gauge symmetry and field content of the string derived model, incl.\ exotic vector-like fields and others which are irrelevant for this analysis, 
		is given in~\cite[Appendix C]{Baur:2022hma}.
}
\end{table}

Generic $\Omega(2)$ compliant super- and K\"ahler potentials have been derived in~\cite{Nilles:2020kgo} and their explicit form can be found in~\cite{Baur:2022hma}. 
For our example model A,
\begin{equation}\label{eq:superpotential}
\begin{split}
   W ~=~  \phi^0 & \left[  \left(\phi^0_\mathrm{u}\,\varphi_\mathrm{u}\right) Y_\mathrm{u}\, H_\mathrm{u}\,\bar{u}\,q + \left(\phi^0_\mathrm{d}\,\varphi_\mathrm{e}\right) Y_\mathrm{d}\, H_\mathrm{d}\, \bar{d}\, q\, 
                     + \left(\phi^0_\mathrm{e}\,\varphi_\mathrm{e}\right) Y_\ell\, H_\mathrm{d}\, \bar{e}\, \ell\right] \\
       &              + \left(\phi^0\varphi_\nu\right) Y_\nu\, H_\mathrm{u}\,\bar\nu\, \ell  + \phi^0_\mathrm{M}\,\varphi_\mathrm{e}\,Y_\mathrm{M}\,\bar\nu\,\bar\nu \,.             
\end{split}                     
\end{equation}\enlargethispage{0.5cm}
Two important empirical observations can be made in this top-down setting: (i) While matter fields can have fractional modular weights, they always combine in such a way
that all Yukawa couplings are modular forms of \textit{integer} weight. (ii) The charge assignments under the eclectic symmetry are \textit{uniquely}
fixed in one-to-one fashion by the modular weight of a field. The latter also holds for all other known top-down constructions, see~\cite{Kikuchi:2021ogn,Baur:2020jwc,Baur:2021mtl,Almumin:2021fbk,Ishiguro:2021ccl}, 
and can be conjectured to be a general feature of top-down models~\cite{Baur:2021bly}.

\section{Sources of eclectic flavor symmetry breaking}
\begin{figure}[t!]
	\centering
	\includegraphics[width=0.4\linewidth]{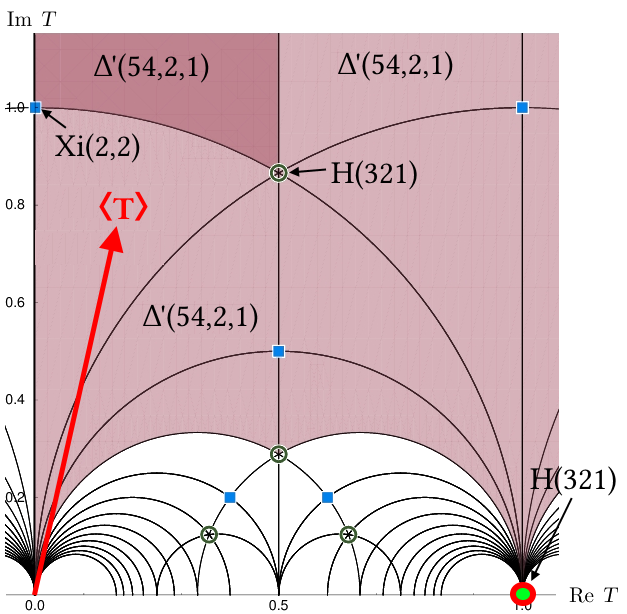}
	\caption{\label{fig:Xi22breaking}
		Residual symmetries of the eclectic flavor symmetry $\Omega(2)$ in dependence of the modulus VEV $\langle T\rangle$ 
		in the bulk of the fundamental domain and at symmetry enhanced special points.
		The point $\vev{T}=\I\infty$ is dual (equivalent by a modular transformation) to the highlighted point $\vev{T}=1$. 
}
\end{figure}

The eclectic flavor symmetry is broken by both, the vacuum expectation value (VEV) of the modulus $\langle T\rangle$ and the VEVs of flavon fields.
This is unlike in virtually all current bottom-up models where either one or the other breaking mechanism is implemented.
Note that all VEVs $\langle T\rangle$ have non-trivial stabilizers in the eclectic symmetry that lead to enhancements
of the residual traditional flavor symmetry beyond what has been previously known in the literature. 
This situation is depicted in figure~\ref{fig:Xi22breaking}.
\begin{figure}
\includegraphics[width=1.0\linewidth]{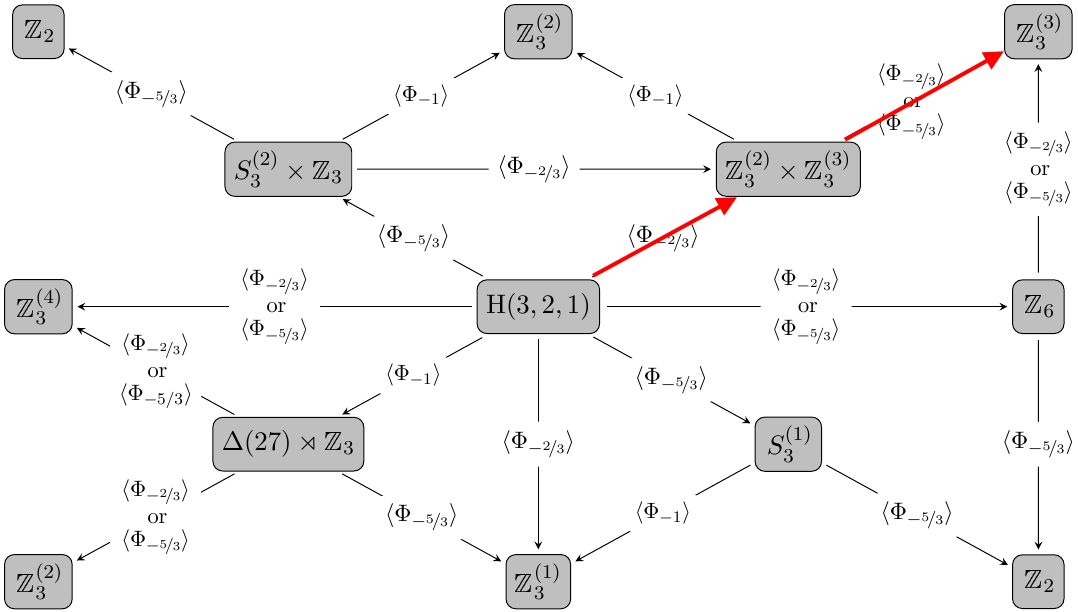}
\caption{\label{fig:H321breaking}
        Possible flavon VEV induced breaking patterns of the linearly realized unified flavor symmetry $H(3,2,1)$ at $\vev{T}=\I\infty$ or $\vev{T}=1$.
        In red we highlight the breaking path we follow in our specific example model and misaligned flavons of the type $\Phi_{\nicefrac{-2}3}$.
}
\end{figure}
For a realistic phenomenology the residual traditional flavor symmetry has to be further broken by the VEVs of flavon fields.
In figure~\ref{fig:H321breaking} we show the possible breaking of the residual flavor symmetry $H(3,2,1)$ at $\langle T\rangle=\I\infty$
by differently (mis-)aligned VEVs of different flavons~\cite{Baur:2021bly}.
Since different residual symmetries are possible for different sectors of the theory, the overall symmetry can be completely broken
even if moduli and VEVs would be stabilized at symmetry enhanced points. In a model with a single modulus and only one type of flavon
we can achieve complete flavor symmetry breaking by misaligning their VEVs slightly away from the symmetry enhanced points.

\begin{figure}
  %\scalebox{0.7}{
	\begin{tikzpicture}[node distance=0.2cm and 2.27cm, rounded corners, >=stealth]
	
	\node[minimum height=22pt, draw, rectangle,fill=lightgray] (Omega2) {$\Omega(2)$ };
	\node[minimum height=22pt, draw, rectangle, right=of Omega2,fill=lightgray] (H321) { $H(3,2,1)$ };
	\node[minimum height=22pt, draw, rectangle, right=of H321,fill=lightgray] (Z3xZ3) { $\;\!\Z{3}^{(2)} \!\times \Z{3}^{(3)}$ };
	\node[minimum height=22pt, draw, rectangle, right = of Z3xZ3,fill=lightgray] (Z3) { $\;\!\Z3^{(3)}$ };
	\node[minimum height=22pt, draw, rectangle, above right=of Z3,fill=lightgray] (empty1) { $~\emptyset~$ };
	\node[minimum height=22pt, draw, rectangle, below right=of Z3,fill=lightgray] (empty2) { $~\emptyset~$ };
	
	\draw[->, >=stealth, shorten >=2pt, shorten <=2pt] (Omega2.east) -- node [fill=white,rectangle,midway,align=center] {\resizebox*{30pt}{!}{$\vev{T}=\I\infty$}} (H321.west);
	\draw[->, >=stealth, shorten >=2pt, shorten <=2pt] (H321.east) -- node [fill=white,rectangle,midway,align=center] {\resizebox*{35pt}{!}{$\vev{\tilde\varphi}=\begin{pmatrix}0\\0\\1\end{pmatrix}$}} (Z3xZ3.west);
	\draw[->, >=stealth, shorten >=2pt, shorten <=2pt] (Z3xZ3.east) -- node [fill=white,rectangle,midway,align=center] {\resizebox*{38pt}{!}{$\vev{\tilde\varphi}=\begin{pmatrix}0\\\lambda_2\\1\end{pmatrix}$}} (Z3.west);
	\draw[->, >=stealth, shorten >=2pt, shorten <=2pt] (Z3.east) -- node [draw=black, fill=white,rectangle,midway,align=center] {\resizebox*{45pt}{!}{$\epsilon=\mathrm{e}^{2\pi\I\vev{T}} \neq 0$}} (empty2.west);
	\draw[->, >=stealth, shorten >=2pt, shorten <=2pt] (Z3.east) -- node [draw=black, fill=white,rectangle,midway,align=center] {\resizebox*{30pt}{!}{$\vev{\tilde\varphi}=\begin{pmatrix}\lambda_1\\\lambda_2\\1\end{pmatrix}$}}(empty1.west);
	\end{tikzpicture}
	\caption{\label{fig:BreakdownPattern}
	Breakdown pattern of the eclectic flavor symmetry $\Omega(2)$ of a $\mathbbm T^2/\Z3$ orbifold model
    triggered by the VEVs of the modulus $T$ and (dimensionless) flavons $\tilde\varphi$. All flavons transform 
    in the $\rep3_2$ representation of $\Delta(54)$, see table~\ref{tab:Representations}.}
%}
\end{figure}
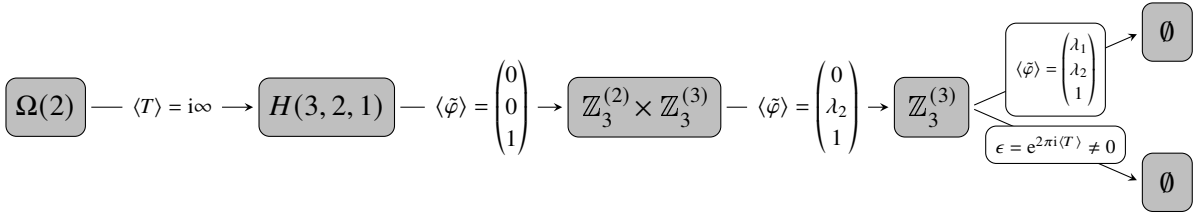

In our specific example model A we follow one specific breaking path that we selected by hand in order to successfully reproduce the 
experimental data. The path is illustrated in red in figure~\ref{fig:H321breaking} and in more detail in figure~\ref{fig:BreakdownPattern}. 
Our model has only flavons of type $\Phi_{\nicefrac{-2}3}$ which transform as $\rep{3}_2$ under the traditional flavor symmetry $\Delta(54)$ as can be inferred from table~\ref{tab:Representations}.
We parametrize the effective (dimensionless) flavon VEVs and the misaligned modulus VEV as
\begin{align}
 &\vev{\tilde\varphi_{\rep{3}_2}}=\left(\lambda_1, \lambda_2, 1\right)\;,& 
 &\epsilon:=\mathrm{e}^{2\pi\I\vev{T}}\;.& 
\end{align}
Exact alignment of the flavon and modulus to the symmetry enhanced point would give rise to a $\Z{3}^{(2)}\times\Z{3}^{(3)}$ residual 
symmetry, with factors
\begin{align}
  &\Z{3}^{(2)}~\subset~G_\mathrm{traditional}& &\text{generated by}& \rho_{\rep{3}_2,\I\infty}(\mathrm{ABA^2})~&=~ \begin{pmatrix}\omega&0&0\\0&\omega^2&0\\0&0&1\end{pmatrix}\;,& \\
  &\Z{3}^{(3)}~\subset~G_\mathrm{modular}& &\text{generated by}& \rho_{\rep{3}_2,\I\infty}(\mathrm{T})~&=~ \begin{pmatrix}\omega^2&0&0\\0&1&0\\0&0&1\end{pmatrix}\;.&
\end{align}
The stepwise breaking of these symmetries, see figure~\ref{fig:BreakdownPattern}, gives rise to technically natural small parameters 
\begin{equation}
 \epsilon, \lambda_1\ll \lambda_2\ll 1 \;,
\end{equation}
which will allow to analytically control our mass and mixing hierarchies.

\section{Mass matrices}
For model A all terms in the superpotential~\eqref{eq:superpotential} have the generic structure
\begin{equation}
\Phi_0 \dots \Phi_0 \,\hat{Y}^{(1)}(T)\, \Phi^{(1)}_{\nicefrac{-2}{3}}\,\Phi^{(2)}_{\nicefrac{-2}{3}}\,\Phi^{(3)}_{\nicefrac{-2}{3}}\;.
\end{equation}
Schematically, this is given by
\begin{equation}
 \text{``singlet flavon(s)} \times \text{modular form} \times \text{\textbf{triplet} matter} \times \text{\textbf{triplet} matter} \times \text{\textbf{triplet} flavon''}\;.
\end{equation}
Hence, the resulting mass matrices for quarks, charged leptons and neutrinos can all be written as~\cite{Nilles:2020kgo,Nilles:2020gvu}
\begin{equation}
\left(\Phi^{(1)}_{\nicefrac{-2}{3}}\right)^\mathrm{T} \; M\left(T, c, \Phi_{\nicefrac{-2}{3}}^{(3)}\right) \;\;\; \Phi^{(2)}_{\nicefrac{-2}{3}}\;,
\end{equation}
with 
\begin{equation}\label{eq:massmatrix}
M\left(T,c,\Phi_{\nicefrac{-2}{3}}^{(3)}\right) ~=~ c~ \begin{pmatrix}
\phantom{-}\hat{Y}_2(T) \,X                 &- \dfrac{\hat{Y}_1(T)}{\sqrt{2}}\, Z&- \dfrac{\hat{Y}_1(T)}{\sqrt{2}} \, Y\\[6pt]
- \dfrac{\hat{Y}_1(T)}{\sqrt{2}} \,Z &\phantom{-}\hat{Y}_2(T) \,Y &- \dfrac{\hat{Y}_1(T)}{\sqrt{2}} \, X\\[6pt]
- \dfrac{\hat{Y}_1(T)}{\sqrt{2}}  \,Y&- \dfrac{\hat{Y}_1(T)}{\sqrt{2}}  \,X&\phantom{-}\hat{Y}_2(T) \,Z
\end{pmatrix}\;.
\end{equation}
Here we have parametrized the effective flavon as $\Phi_{\nicefrac{-2}{3}}^{(3)}\equiv\left(X,Y,Z\right)$, 
and used the modular form 
\begin{equation}
\hat{Y}^{(1)}(T)\equiv
\begin{pmatrix} \hat{Y}_1(T) \\ \hat{Y}_2(T) \end{pmatrix} \equiv \frac{1}{\eta(T)}
\begin{pmatrix}
-3\sqrt{2}\,\eta^3(3\,T)\\
3\eta^3(3\,T) + \eta^3(T/3)
\end{pmatrix},
\end{equation}
where $\eta$ is the Dedekind function. In the vicinity of the symmetry enhanced points discussed above
the mass matrices all take the form 
%\end{equation}
\begin{equation}
M\left(\vev T,\Lambda,\vev{\tilde{\varphi}}\right)=\Lambda\,
\begin{pmatrix}
\lambda_1                    & 3\,\epsilon^{1/3}            & 3\,\lambda_2\,\epsilon^{1/3}\\
3\,\epsilon^{1/3}            & \lambda_2                    & 3\,\lambda_1\,\epsilon^{1/3}\\
3\,\lambda_2\,\epsilon^{1/3} & 3\,\lambda_1\,\epsilon^{1/3} & 1
\end{pmatrix}
\,+\,\mathcal{O}(\epsilon)\;.
\end{equation}
The exact values of the parameters $\lambda_{1,2}$ and the overall scale $\Lambda$ are different for the different sectors, but this shows the analytic control 
over the hierarchical entries in the mass matrices.

\section{Numerical analysis: fit to data}
To give a proof of existence of functioning top-down models we fit the parameters of our model
to the observed data in lepton and quark sectors. As input, we take the mass ratios and $1\sigma$ errors for charged lepton masses, quark masses and quark mixings at the GUT scale, see e.g.~\cite{Antusch:2013jca}, 
assuming RGE running with benchmark parameters $\tan\beta=10$, $M_\mathrm{SUSY}=10$\,TeV, and $\bar\eta_b=0.09375$, as is common practice in bottom up constructions~\cite{Feruglio:2017spp,Chen:2021zty,Ding:2021zbg}.
The data on the lepton mixing was taken from the global analysis $\mathrm{NuFIT v}5.1$~\cite{Esteban:2020cvm} including the full dependence of $\Delta\chi^2$ profiles.
While the leptonic mixing parameters are given at the low scale, the correction from RGE running in our type-I seesaw scenario is expected to be smaller 
than the experimental errors, see e.g.~\cite{Antusch:2003kp}, such that we ignore the effect of running for those.

We define a $\chi^2$ function 
 \begin{equation}
\chi^2(x) := \sum_i \dfrac{\mu_{i,\mathrm{exp}} - \mu_{i,\mathrm{model}}(x)}{\sigma_i}\;,
\end{equation}
where $\mu_\mathrm{exp}$ and $\sigma$ are experimental best-fit value and $1\sigma$ error,
while $\mu_\mathrm{model}$ is the model prediction.
In order to fix the free parameters of our model we numerically minimize $\chi^2$ using~\texttt{lmfit}~\cite{lmfit}.
Subsequently we explore each minimum with the Markov-Chain-Monte-Carlo sampler~\texttt{emcee}~\cite{emcee}.

\subsection{Lepton sector}
\begin{figure}[t]
	\centering
	\includegraphics[width=0.96\linewidth]{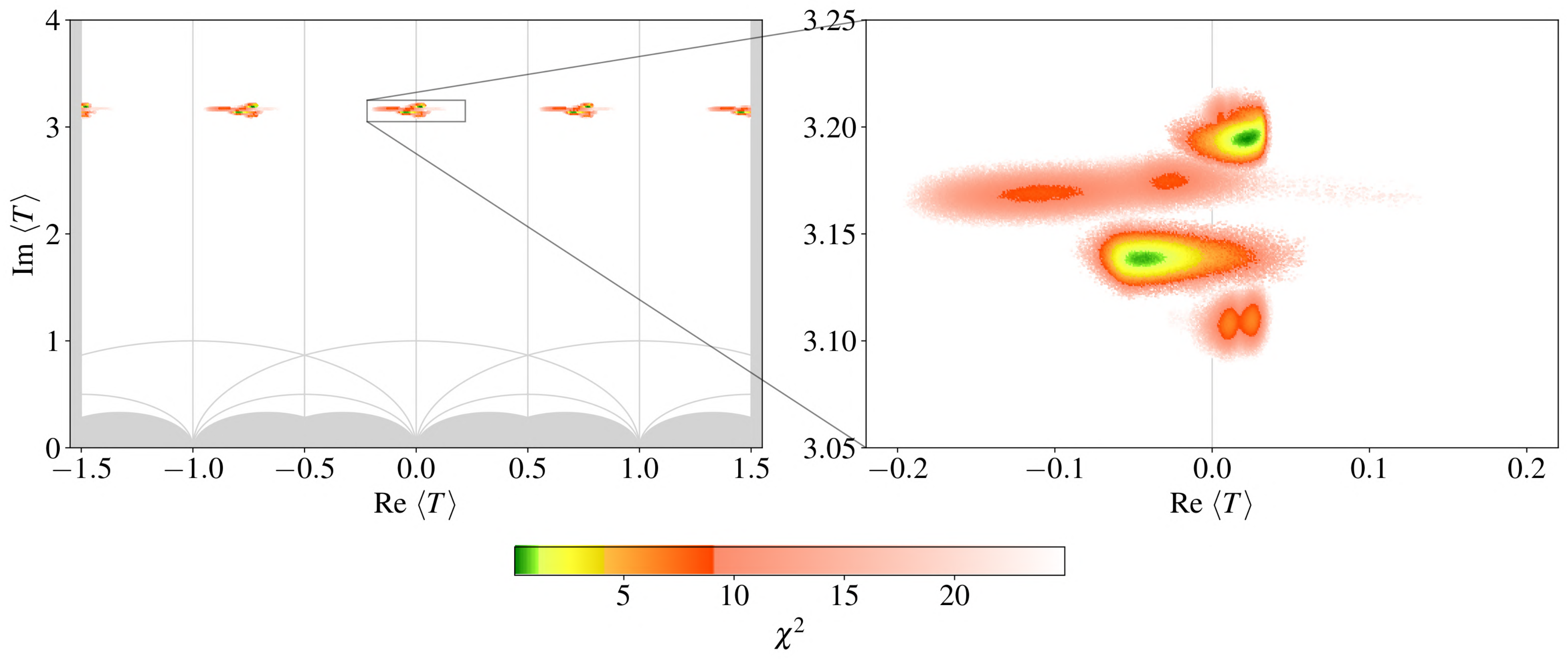}
	\caption{Regions in the fundamental domain of $\Gamma(3)$ that yield fits for $\vev{T}$ with $\chi^2\leq25$. 
		The green, yellow, and orange regions show the $1$, $2$, and $3\sigma$ confidence intervals. 
		The best-fit value of the model lies in the upper right green region. 
		\label{fig:modulispace}}
\end{figure}

\begin{figure}
\centering
\begin{minipage}[c]{0.5\linewidth}
 \includegraphics[width=1.0\linewidth]{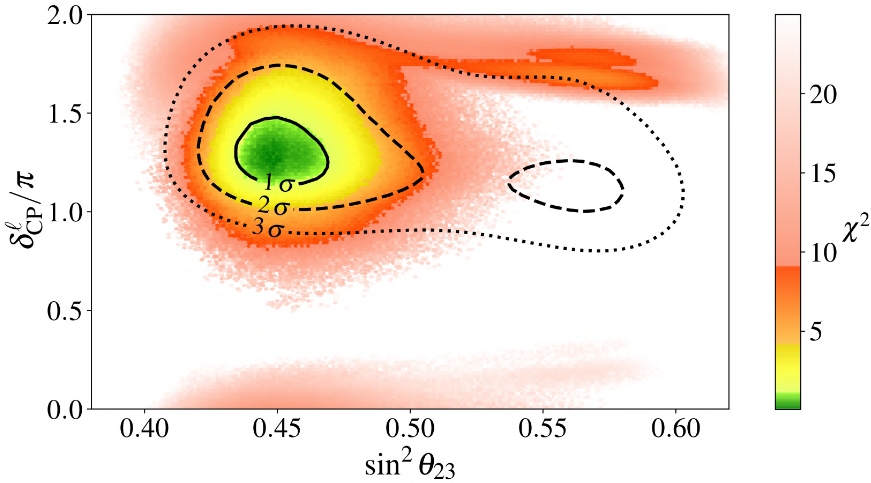}
\end{minipage}%
\begin{minipage}[c]{0.50\linewidth}
 \hspace{0.25cm}\includegraphics[width=0.97\linewidth]{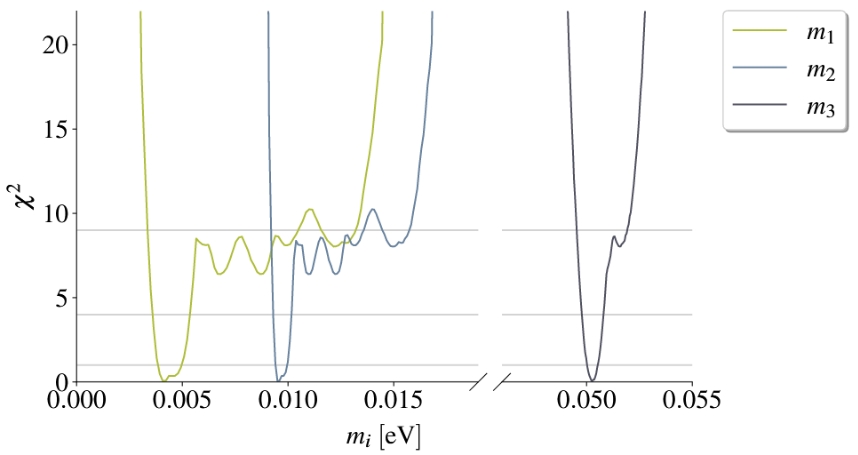}
\end{minipage} \\
\begin{minipage}[c]{0.50\linewidth}
 \includegraphics[width=0.95\linewidth]{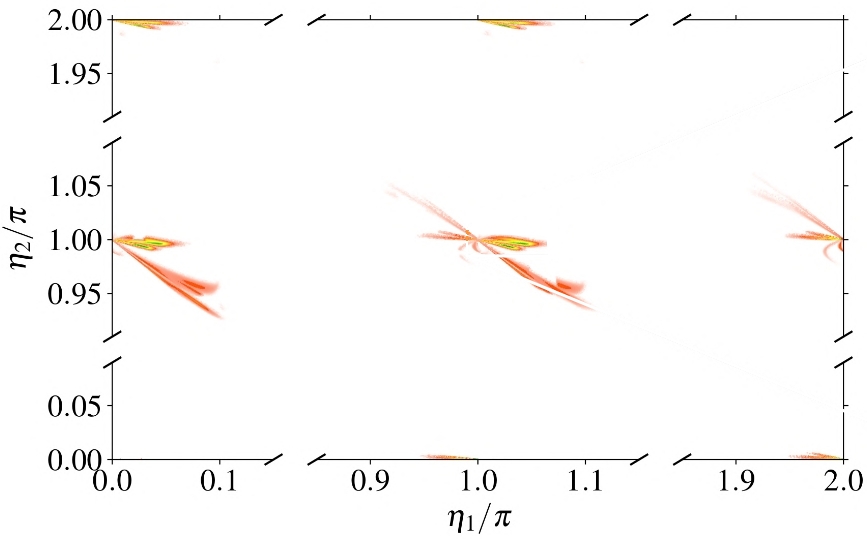}
\end{minipage}%
\begin{minipage}[c]{0.50\linewidth}
 \includegraphics[width=0.94\linewidth]{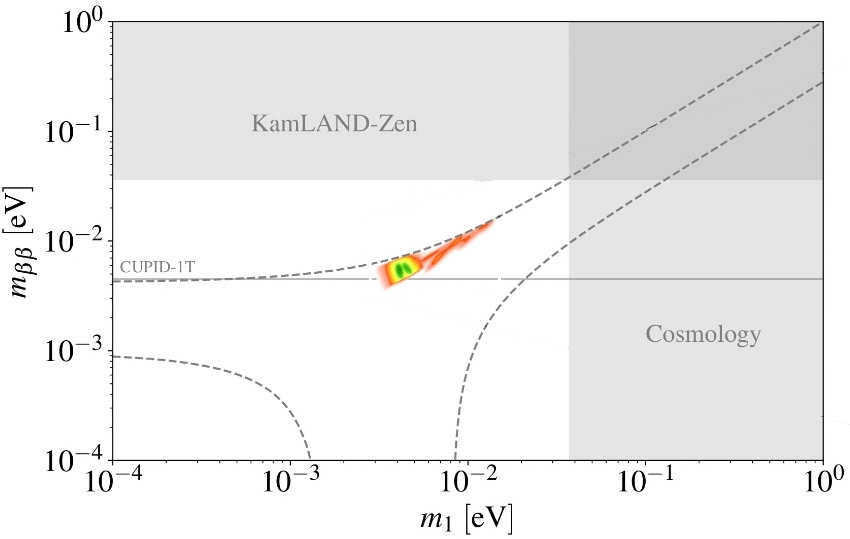}
\end{minipage}% 
\caption{\label{fig:neutrinomasses}
Best fit points of our model and projections on experimentally observable parameters of the lepton sector, see text for details.}
\end{figure}

For the fit to the lepton sector there are effectively only $7$ parameters given by
\begin{equation}
x~=~ \left\{\re\,\vev T,\, \im\,\vev T,\, 
\vev{\tilde\varphi_{\mathrm{e},1}},\,
\vev{\tilde\varphi_{\mathrm{e},2}},\,
\vev{\tilde\varphi_{\nu,1}},\,
\vev{\tilde\varphi_{\nu,2}},\,
\Lambda_\nu \right\}\,.
\end{equation}
The best-fit results are shown in table~\ref{tab:LeptonFitParameters}. 
The fit is bimodal as clearly seen in figure~\ref{fig:modulispace}, which also shows the best-fit point for the 
expectation value of the modulus. The corresponding values of the experimental parameters and their best-fit values 
in our model are collected in table~\ref{tab:FitLeptons}. The fit to the data is only successfully possible if
\begin{enumerate}
 \item atmospheric mixing lies in the lower octant $\theta_{23}^{\ell}<45^\circ$,
 \item neutrino masses obey a normal ordering with masses at $1\sigma$ predicted to be $3.9~\mathrm{meV} < m_1 < 4.9~\mathrm{meV}$, $9.5~\mathrm{meV} < m_2 < 9.9~\mathrm{meV}$, $50.1~\mathrm{meV} < m_3 < 50.5~\mathrm{meV}$, and,
 \item the Majorana phases are close to the CP conserving values $\eta_{1,2}\approx\pi$.
\end{enumerate}
These can be considered predictions of this scenario. The corresponding posteriors are shown in figure~\ref{fig:neutrinomasses}
together also with the allowed effective neutrino mass for $0\nu\beta\beta$-decay on the lower right. 
Gray-shaded areas are excluded by KamLAND-Zen~\cite{KamLAND-Zen:2022tow} or cosmology~\cite{GAMBITCosmologyWorkgroup:2020rmf,Planck:2018vyg}.
Future generations of $0\nu\beta\beta$-decay experiments such as CUPD-1T~\cite{CUPID:2022wpt} are expected to probe the available parameter space.
The model does not constrain the \CP violating phase $\delta_{\mathrm{\CP}}^\ell$ better than the combined experimental information.
\begin{table}[t]
	\centering\vspace{-0.5cm}
\resizebox{\textwidth}{!}{
\begin{tabular}{llclc}
			\toprule
			&\multicolumn{2}{c}{right green region} & \multicolumn{2}{c}{left green region}\\\cmidrule(r{4pt}l{4pt}){2-3}\cmidrule(l{4pt}r{4pt}){4-5}
			parameter~~                          & best-fit value~        & $1\sigma$ interval & best-fit value~        & $1\sigma$ interval \\\midrule
			$\re\, \vev T$                       & $\phantom{-}0.02279$    & $0.01345 \rightarrow 0.03087$ & $-0.04283$    & $-0.05416 \rightarrow -0.02926$\\
			$\im\, \vev T$                       & $\phantom{-}3.195$      & $3.191 \rightarrow 3.199$ & $\phantom{-}3.139$      & $3.135 \rightarrow 3.142$\\
			$\vev{\tilde\varphi_{\mathrm{e,1}}}$ & $-4.069 \cdot 10^{-5}$ & $-4.321 \cdot 10^{-5}\rightarrow -3.947 \cdot 10^{-5}$ & $\phantom{-}2.311 \cdot 10^{-5}$ & $2.196 \cdot 10^{-5}\rightarrow 2.414 \cdot 10^{-5}$\\
			$\vev{\tilde\varphi_{\mathrm{e,2}}}$ & $\phantom{-}0.05833$    & $0.05793 \rightarrow 0.05876$ & $\phantom{-}0.05826$    & $0.05792 \rightarrow 0.05863$\\
			$\vev{\tilde\varphi_{\mathrm{\nu,1}}}$ & $\phantom{-}0.001224$ & $0.001201 \rightarrow 0.001248$  & $-0.001274$ & $-0.001304 \rightarrow -0.001248$ \\ 
			$\vev{\tilde\varphi_{\mathrm{\nu,2}}}$ & $-0.9857$             & $-1.0128 \rightarrow -0.9408$ & $\phantom{-}0.9829$             & $0.9433 \rightarrow 1.0122$\\
			$\Lambda_\nu~[\mathrm{eV}]$          & $\phantom{-}0.05629$    & $0.05442 \rightarrow 0.05888$ & $\phantom{-}0.05591$    & $0.05408 \rightarrow 0.05850$ \\\midrule
			$\,\chi^2$                           & $\phantom{-}0.08$     && $\phantom{-}0.45$ & \\
			\bottomrule
		\end{tabular}}
	\caption{Best-fit values for the free model parameters in the lepton sector and their corresponding $1\sigma$ intervals for the two best-fit regions (green) also visible in figure~\ref{fig:modulispace}.
	\label{tab:LeptonFitParameters}}
\end{table}

\begin{table}
	\centering
	\resizebox{\textwidth}{!}{
		\begin{tabular}{lrrrrrr}
			\toprule
			& \multicolumn{3}{c}{model} &\multicolumn{3}{c}{experiment}\\\cmidrule(r{4pt}l{4pt}){2-4}\cmidrule(l{4pt}r{4pt}){5-7}
			observable & best fit & $1\sigma$ interval & $3\sigma$ interval & best fit & $1\sigma$ interval & $3\sigma$ interval\\\midrule
			$m_\mathrm{e}/m_\mu$ & $0.00473$ & $0.00470\rightarrow0.00477$ & $0.00462\rightarrow0.00485$ & $0.00474$ & $0.00470\rightarrow0.00478$ & $0.00462\rightarrow0.00486$\\
			$m_\mu/m_\tau$ & $0.0586$ & $0.0581\rightarrow0.0590$ & $0.0572\rightarrow0.0600$ & $0.0586$ & $0.0581\rightarrow0.0590$ & $0.0572\rightarrow0.0600$\\\midrule
			$\sin^2\theta_{12}$ & $0.303$ & $0.294\rightarrow0.315$ & $0.275\rightarrow0.335$ & $0.304$ & $0.292\rightarrow0.316$ & $0.269\rightarrow0.343$ \\
			$\sin^2\theta_{13}$ & $0.02254$ & $0.02189\rightarrow0.02304$ & $0.02065\rightarrow0.02424$ & $0.02246$ & $0.02184\rightarrow0.02308$  & $0.02060\rightarrow0.02435$\\
			$\sin^2\theta_{23}$ & $0.449$ & $0.436\rightarrow0.468$ & $0.414\rightarrow0.593$ & $0.450$ & $0.434 \rightarrow 0.469$ & $0.408 \rightarrow 0.603$ \\\midrule
			$\delta^\ell_\CP/\pi$ & $1.28$ & $1.15\rightarrow1.47$ & $0.81\rightarrow1.94$ & $1.28$ & $1.14 \rightarrow 1.48$ & $0.80 \rightarrow 1.94$ \\
			$\eta_1/\pi \mod 1$ & $0.029$ & $0.018\rightarrow0.048$ & $-0.031\rightarrow0.090$ & - & - & - \\
			$\eta_2/\pi \mod 1$ & $0.994$ & $0.992\rightarrow0.998$ & $0.935\rightarrow1.004$ & - & - & - \\
			$J_\CP$ & $-0.026$ & $-0.033\rightarrow-0.015$ & $-0.035\rightarrow0.019$ & $-0.026$ & $-0.033\rightarrow-0.016$ & $-0.033\rightarrow0.000$ \\
			$J_\CP^\mathrm{max}$ & $0.0335$ & $0.0330\rightarrow0.0341$ & $0.0318\rightarrow0.0352$ & $0.0336$ & $0.0329\rightarrow0.0341$ & $0.0317\rightarrow0.0353$ \\\midrule
			$\Delta m_{21}^2/10^{-5}~[\mathrm{eV}^2]$ & $7.39$ & $7.35\rightarrow7.49$ & $7.21\rightarrow7.65$ & $7.42$ & $ 7.22\rightarrow 7.63$ & $6.82 \rightarrow 8.04$ \\
			$\Delta m_{31}^2/10^{-3}~[\mathrm{eV}^2]$ & $2.508$ & $2.488\rightarrow2.534$ & $2.437\rightarrow2.587$ & $2.521$ & $ 2.483\rightarrow 2.537$ & $2.430 \rightarrow 2.593$ \\
			$m_1~[\mathrm{eV}]$ & $0.0042$ & $0.0039\rightarrow0.0049$ & $0.0034\rightarrow0.0131$ & $<0.037$ & - & - \\
			$m_2~[\mathrm{eV}]$ & $0.0095$ & $0.0095\rightarrow0.0099$ & $0.0092\rightarrow0.0157$ & - & - & - \\
			$m_3~[\mathrm{eV}]$ & $0.0504$ & $0.0501\rightarrow0.0505$ & $0.0496\rightarrow0.0519$ & - & - & - \\
			$\sum_i m_i~[\mathrm{eV}]$ & $0.0641$ & $0.0636\rightarrow0.0652$ & $0.0628\rightarrow0.0806$ & $<0.120$ & - & - \\
			$m_{\beta\beta}~[\mathrm{eV}]$ & $0.0055$ & $0.0045\rightarrow0.0064$ & $0.0040\rightarrow0.0145$ & $<0.036$ & - & - \\
			$m_{\beta}~[\mathrm{eV}]$ & $0.0099$ & $0.0097\rightarrow0.0102$ & $0.0094\rightarrow0.0159$ & $<0.8$ & - & - \\\midrule
			$\chi^2$ & $0.08$ & & & & & \\
			\bottomrule
		\end{tabular}
	}
	\caption{Lepton sector best-fit values of our model compared to the 
		experimental data.
		\label{tab:FitLeptons}}
\end{table}

\subsection{Simultaneous fit to the quark sector and importance of K\"ahler corrections}
Next, we extend our fit to the quark sector. As visibile from  the superpotential~\eqref{eq:superpotential}, the up-type quark Yukawa couplings include an additional flavon triplet $\varphi_\mathrm{u}$ while 
the down-type Yukawa couplings share the flavon triplet of the charged leptons $\varphi_\mathrm{e}$ (this is a specific feature of this model~A and cannot be changed 
as it is determined by the underlying string theory). 
Since the structure of the mass matrices is tightly fixed, see equation~\eqref{eq:massmatrix}, this implies that at leading order in the EFT, masses of charged leptons and down-type quarks would only differ by their overall scale, which contradicts experimental observation. 
However, this is only a leading order statement, and the superpotential~\eqref{eq:superpotential}, in principle, is subject
to corrections originating from a non-canonical K\"ahler potential.
Including these corrections allows us to obtain a successful fit to quark and lepton sector simultaneously.

K\"ahler corrections have usually not been taken into account in bottom-up constructions 
even though they are unconstrained there~\cite{Chen:2019ewa} and, therefore, potentially destabilize predictions.
Unlike in pure modular flavor theories, the traditional flavor symmetry present in the full eclectic picture
allows one to keep control of the K\"ahler potential. In particular, $K$ is canonical at leading order~\cite{Nilles:2020kgo}.
As discussed in detail in~\cite{Baur:2022hma}, there are considerable off-diagonal corrections to the K\"ahler metric
at next-to-leading order if flavons develop VEVs that break the traditional flavor symmetry.

While K\"ahler corrections, in principle, can affect both lepton and quark sectors,
we only take into account the K\"ahler corrections to quarks, for simplicity, and ignore the corrections to the lepton sector.
This is not worse than the common assumption of total absence of these corrections in bottom-up constructions.
In any case, it is conceivable that the inclusion of additional parameters would only make our fit better, not worse.
For the quark sector of model A the corrections are essential and must be included to obtain a good fit.

The following discussion of K\"ahler corrections is specific to model A. 
Schematically, the corrections at leading order (LO) and next-to-leading order (NLO) are given by~\cite{Baur:2022hma}
\begin{align}
  K_\mathrm{LO} &\supset -\log(-\I T+\I \bar T) + \sum_{\Phi}\left[(-\I T+ \I\bar T)^{\nicefrac{-2}3}
        + (-\I T+ \I\bar T)^{\nicefrac13}|\hat Y^{(1)}(T)|^2\right]|\Phi|^2\;,& \\
K_\mathrm{NLO} &\supset \sum_{\Psi,\varphi} \left[(-\I T+\I \bar T)^{\nicefrac{-4}3}\sum_a |\Psi\varphi|^2_{\rep1,a}
                          +(-\I T+\I\bar T)^{\nicefrac{-1}3}\sum_a |\hat Y^{(1)}(T)\Psi\varphi|^2_{\rep1,a}\right]\;.&
\end{align}
For a given quark flavor $f=\left\{\mathrm{u,d,q}\right\}$,
\begin{equation}
 K^{(f)}_{ij} \approx \chi^{(f)} ~\left[\delta_{ij}+ \lambda_{\varphi_\mathrm{eff}}^{(f)}\, \left(A_{ij}^{(f)}+ \kappa_{\varphi_\mathrm{eff}}^{(f)}\, B_{ij}^{(f)}\right)\right]\;,
\end{equation}
with flavor space structures $A=A(\varphi,T)$ and $B=B(\varphi,T)$ that are fixed by group theory
but depend on \textit{all} flavon fields. 
We can define ``effective flavons'' such that 
\begin{align}
\sum_\varphi \lambda^{(f)}_\varphi \, A_{ij}(\varphi) ~&=:~ \lambda^{(f)}_{\varphi_{\mathrm{eff}}} \, A_{ij}(\tilde\varphi_{\mathrm{eff}}^{(A,f)})& 
&\equiv\lambda^{(f)}_{\varphi_{\mathrm{eff}}} \, A^{(f)}_{ij}\;,&\\
\sum_\varphi \lambda^{(f)}_\varphi \, \kappa^{(f)}_\varphi \, B_{ij}(\varphi) ~&=:~ \lambda^{(f)}_{\varphi_{\mathrm{eff}}} \, \kappa^{(f)}_{\varphi_{\mathrm{eff}}} \, B_{ij}(\tilde\varphi_{\mathrm{eff}}^{(B,f)})&
&\equiv\lambda^{(f)}_{\varphi_{\mathrm{eff}}} \, \kappa^{(f)}_{\varphi_{\mathrm{eff}}} \, B^{(f)}_{ij}\;.
\end{align}
The tilde here is used once we took the scale out of the flavon directions
\begin{equation}
 \tilde\varphi_{\mathrm{eff}}^{(A,B)} := \varphi_{\mathrm{eff}}^{(A,B)}/\Lambda_{\varphi_{\mathrm{eff}}^{(A,B)}} 
 \quad\text{such that}\quad 
\tilde\varphi_{\mathrm{eff}}^{(A,B)} ~:=~ \left(\tilde\varphi_{\mathrm{eff},1}^{(A,B)},\,\tilde\varphi_{\mathrm{eff},2}^{(A,B)},\,1\right)^\mathrm{T}\,.
\end{equation}
Finally, we can define the parameters
\begin{align}
\alpha_i^{(f)} &:= \sqrt{\lambda_{\varphi_{\mathrm{eff}}}^{(f)}} \, \vev{\tilde{\varphi}_{\mathrm{eff},i}^{(A,f)}}\;,&
\beta_i^{(f)} &:= \sqrt{\lambda_{\varphi_{\mathrm{eff}}}^{(f)}} \, \vev{\tilde{\varphi}_{\mathrm{eff},i}^{(B,f)}}\;,&
\end{align}
and one can show that 
\begin{align}
\lambda_{\varphi_{\mathrm{eff}}}^{(f)} A_{ij}^{(f)} &= \alpha_i^{(f)}\,\alpha_j^{(f)}\;,&
\lambda_{\varphi_{\mathrm{eff}}}^{(f)} B_{ij}^{(f)} &\approx \beta_i^{(f)}\,\beta_j^{(f)}\;.&
\end{align}
%and $\lambda_{\varphi_{\mathrm{eff}}}^{(a)} B_{ij}^{(a)}$ is also quadratic in $\alpha^{(a)}$ up to $\mathcal{O}(1)$ terms.
The parameters $\alpha_i^{f}$ and $\beta_i^{f}$ represent a good measure of the size 
of the K\"ahler corrections.

Altogether, the parameters of the quark sector are given by the components of the up-type flavon triplet
\begin{equation}
\label{eq:uflavon}
   \vev{\tilde\varphi_\mathrm{u}} ~=~ \Big(
   \vev{\tilde\varphi_{\mathrm{u},1}} \,\exp\;\!\!\left(\I\vev{\vartheta_{\mathrm{u},1}}\right), 
   \vev{\tilde\varphi_{\mathrm{u},2}} \,\exp\;\!\!\left(\I\vev{\vartheta_{\mathrm{u},2}}\right), 
   1\Big)\;,
\end{equation}
and the K\"ahler corrections additionally introduce 9 parameters $\alpha_i^f$, 9 $\beta_i^f$ and 3 $\kappa_{\varphi_\mathrm{eff}}^{f}$. 
To reduce the number of parameters we impose the constraints:
\begin{itemize}
  \item $\kappa_{\varphi_\mathrm{eff}}^{f}=1$ for all $f\in\{\mathrm{u,d,q}\}$,
  \item $\alpha_i^{f} = \beta_i^{f}$ for all $f$ and $i\in\{1,2,3\}$, and
  \item all $\alpha_i^{f}\in\mathbb{R}$.
\end{itemize}
We recall that the philosophy here is not to scan the full parameter space but to identify
a region in the parameter space that agrees with realistic phenomenology in the first place.
Including the constraints we arrive at a total of 13 quark parameters that we include in our fit 
of both, leptons and quarks. 

The resulting best-fit values are collected in table~\ref{tab:SimultaneousFit}.
The magnitude of all required K\"ahler corrections satisfies $\alpha^f_i<1$.
The modulus VEV \vev{T} and the VEVs of the charged lepton and neutrino flavons $\vev{\tilde\varphi_{\mathrm{e},i}}$ and 
$\vev{\tilde\varphi_{\nu,i}}$ stay at the values obtained in the exclusive lepton fit, see table~\ref{tab:FitLeptons}.
Table~\ref{tab:SimultaneousFitObservables} shows our best fit compared to the experimental input values 
of quark and lepton parameters. The best fit function to all fermion mass ratios, mixing angles and 
\CP phases yields $\chi^2=0.11$.  Even though the quark sector fit is not predictive 
we have fulfilled our goal to show that the eclectic scenario arising from a string compactification 
can fit the observed data well.
\begin{table}[!t!]
	\vspace*{10pt}
	\begin{minipage}{0.25\textwidth}%
		\centering
		\subfloat[(a)][]{
			\label{tab:SimultaneousFitParameters}
			\centering 
			\scalebox{0.8}{
				\begin{tabular}{cr|l}
					\toprule
					\multicolumn{2}{r|}{parameter} & best-fit value \\\midrule
					&$\im\, \vev T$ & $\phantom{-}3.195$ \\
					&$\re\,\vev T$ & $\phantom{-}0.02279$ \\
					&$\vev{\tilde{\varphi}_{\mathrm{u},1}}$ & $\phantom{-}2.0332\cdot10^{-4}$ \\
					&$\vev{\vartheta_{\mathrm{u},1}}$ & $\phantom{-}1.6481$ \\
					&$\vev{\tilde{\varphi}_{\mathrm{u},2}}$ & $\phantom{-}6.3011\cdot10^{-2}$ \\
					&$\vev{\vartheta_{\mathrm{u},2}}$ & $-1.5983$ \\
					&$\vev{\tilde{\varphi}_{\mathrm{e},1}}$ & $-4.069\cdot 10^{-5}$ \\
					&$\vev{\tilde{\varphi}_{\mathrm{e},2}}$ & $\phantom{-}5.833\cdot10^{-2}$ \\
					&$\vev{\tilde{\varphi}_{\nu,1}}$ & $\phantom{-}1.224\cdot10^{-3}$ \\
					&$\vev{\tilde{\varphi}_{\nu,2}}$ & $-0.9857$ \\
					\multirow{-11}{10pt}{\rotatebox{90}{superpotential}}&$\Lambda_\nu~[\mathrm{eV}]$   & $\phantom{-}0.05629$\\\midrule
					&$\alpha_1^{\mathrm{u}}$ & $-0.94917$ \\
					&$\alpha_2^{\mathrm{u}}$ & $\phantom{-}0.0016906$ \\
					&$\alpha_3^{\mathrm{u}}$ & $\phantom{-}0.31472$ \\
					&$\alpha_1^{\mathrm{d}}$ & $\phantom{-}0.95067$ \\
					&$\alpha_2^{\mathrm{d}}$ & $\phantom{-}0.0077533$ \\
					&$\alpha_3^{\mathrm{d}}$ & $\phantom{-}0.30283$ \\
					&$\alpha_1^{\mathrm{q}}$ & $-0.96952$ \\
					&$\alpha_2^{\mathrm{q}}$ & $-0.20501$ \\
					\multirow{-9}{10pt}{\rotatebox{90}{K\"ahler potential}}&$\alpha_3^{\mathrm{q}}$ & $\phantom{-}0.041643$ \\
					\bottomrule
			\end{tabular}}
		}
		\vspace*{37mm}
	\end{minipage}%
	\hspace*{28pt}
	\begin{minipage}{0.7\textwidth}%
		\centering
		\subfloat[(b)][]{
			\label{tab:SimultaneousFitObservables}
			\centering 
			\scalebox{0.8}{
				\begin{tabular}{cr|llc}
					\toprule
					\multicolumn{2}{r|}{observable} & model best fit & exp. best fit & exp. $1\sigma$ interval \\\midrule
					&$m_\mathrm{u}/m_\mathrm{c}$ & $\phantom{-}0.00193$ & $\phantom{-}0.00193$ & $0.00133 \rightarrow 0.00253$ \\
					&$m_\mathrm{c}/m_\mathrm{t}$ & $\phantom{-}0.00280$ & $\phantom{-}0.00282$ & $0.00270 \rightarrow 0.00294$\\
					&$m_\mathrm{d}/m_\mathrm{s}$ & $\phantom{-}0.0505$ & $\phantom{-}0.0505$ & $0.0443 \rightarrow 0.0567$\\
					&$m_\mathrm{s}/m_\mathrm{b}$ & $\phantom{-}0.0182$ & $\phantom{-}0.0182$ & $0.0172 \rightarrow 0.0192$ \\\cmidrule{2-5}
					&$\vartheta_{12}~[\mathrm{deg}]$ & $\,\;\!13.03$ & $\,\;\!13.03$ & $12.98 \rightarrow 13.07$\\
					&$\vartheta_{13}~[\mathrm{deg}]$ & $\phantom{-}0.200$ & $\phantom{-}0.200$ & $0.193 \rightarrow 0.207$ \\
					&$\vartheta_{23}~[\mathrm{deg}]$ & $\phantom{-}2.30$ & $\phantom{-}2.30$ & $2.26 \rightarrow 2.34$\\
					\multirow{-8}{10pt}{\rotatebox{90}{quark sector}}	&$\delta_\CP^\mathrm{q}~[\mathrm{deg}]$ & $\,\;\!69.2$ & $\,\;\!69.2$ & $66.1 \rightarrow 72.3$\\\midrule		
					&$m_\mathrm{e}/m_\mu$ & $\phantom{-}0.00473$ & $\phantom{-}0.00474$ & $0.00470\rightarrow0.00478$ \\
					&$m_\mu/m_\tau$ & $\phantom{-}0.0586$ & $\phantom{-}0.0586$ & $0.0581\rightarrow0.0590$ \\\cmidrule{2-5}
					&$\sin^2\theta_{12}$ & $\phantom{-}0.303$ & $\phantom{-}0.304$ & $0.292\rightarrow0.316$ \\
					&$\sin^2\theta_{13}$ & $\phantom{-}0.0225$ & $\phantom{-}0.0225$ & $0.0218\rightarrow0.0231$ \\
					&$\sin^2\theta_{23}$ & $\phantom{-}0.449$ & $\phantom{-}0.450$ & $0.434 \rightarrow 0.469$ \\\cmidrule{2-5}
					&$\delta_\CP^\ell/\pi$ & $\phantom{-}1.28$ & $\phantom{-}1.28$ & $1.14 \rightarrow 1.48$ \\
					&$\eta_1/\pi$ & $\phantom{-}0.029$ & ~~~\,~- & - \\
					&$\eta_2/\pi$ & $\phantom{-}0.994$ & ~~~\,~- & - \\
					&$J_\CP$ & $-0.026$ & $-0.026$ & $-0.033\rightarrow-0.016$ \\
					&$J_\CP^\mathrm{max}$ & $\phantom{-}0.0335$ & $\phantom{-}0.0336$ & $0.0329\rightarrow0.0341$ \\\cmidrule{2-5}
					&$\Delta m_{21}^2/10^{-5}~[\mathrm{eV}^2]$ & $\phantom{-}7.39$ & $\phantom{-}7.42$ & $ 7.22\rightarrow 7.63$ \\
					&$\Delta m_{31}^2/10^{-3}~[\mathrm{eV}^2]$ & $\phantom{-}2.521$ & $\phantom{-}2.510$ & $ 2.483\rightarrow 2.537$ \\
					&$m_1~[\mathrm{eV}]$ & $\phantom{-}0.0042$ & $\!<\!0.037$ & - \\
					&$m_2~[\mathrm{eV}]$ & $\phantom{-}0.0095$ & ~~~\,~- & -  \\
					&$m_3~[\mathrm{eV}]$ & $\phantom{-}0.0504$ & ~~~\,~- & - \\
					&$\sum_i m_i~[\mathrm{eV}]$ & $\phantom{-}0.0641$ & $\!<\!0.120$ & -  \\
					&$m_{\beta\beta}~[\mathrm{eV}]$ & $\phantom{-}0.0055$ & $\!<\!0.036$ & - \\
					\multirow{-18}{10pt}{\rotatebox{90}{lepton sector}}&$m_{\beta}~[\mathrm{eV}]$ & $\phantom{-}0.0099$ & $\!<\!0.8$ & - \\\midrule
					& $\chi^2$ & $\phantom{-}0.11$ & & \\
					\bottomrule	
				\end{tabular}
			}
		}
	\end{minipage}%
	\caption{\label{tab:SimultaneousFit}
	Best fit results of our model in a simultaneous fit to quark and lepton sectors. 
	(a) Best-fit values for the free model parameters.
	The lepton sector and modulus parameters agree with the one of the exclusively leptonic fit shown in table~\ref{tab:LeptonFitParameters}. 
	(b) Best fit points of our model as compared to experimentally determined parameters.}
\end{table}

\section{Possible lessons for consistent bottom-up model building}
Given an explicit example of a complete top-down model, we make some empirical observations that might be taken as 
useful guidelines for bottom-up constructions:
(i) Neither modular nor traditional flavor symmetries arise alone. They arise as mutualy overlapping parts of the full eclectic flavor symmetry, including also
\CP-type and R symmetries, see~\eqref{eq:eclectic}. (ii) Modular weights of matter fields are fractional, while modular weights of (Yukawa) couplings are integer.
(iii) Modular weights are $1:1$ ``locked'' to other flavor symmetry representations. This holds true for all known top-down constructions~\cite{Kikuchi:2021ogn,Baur:2020jwc,Baur:2021mtl,Almumin:2021fbk,Ishiguro:2021ccl} and might be conjectured to be 
a general feature of top-down models~\cite{Baur:2021bly}. (iv) Different sectors of the theory may have different
moduli and/or different residual symmetries allowing for what has been called ``local flavor unification''~\cite{Baur:2019kwi}.
If all these features would indeed be confirmed on other UV complete top-down constructions one may anticipate that in a modern
language, the modular flavor ``swampland'' may be much bigger than anticipated.

\section{Important open problems}
We stress directions in which our discussion can be generalized and important open problems.
The additional compact dimensions in string theory may give rise to non-trivially interlinked extra tori with additional moduli, giving rise to 
metaplectic groups and their corresponding flavor symmetries~\cite{Ding:2020zxw,Ding:2021iqp,Nilles:2021glx}.
Also, it would be important to investigate other consistent string configurations for possibly realistic eclectic 
flavor scenarios to get a grasp of the \mbox{``size of the realistic `landscape' ''.} 

Note also that for the sake of our discussion we have taken VEVs of the flavon fields as well as the size of the K\"ahler 
corrections as free parameters of our model. However, in a full string model the computation 
of the flavon potential and the dynamic stabilization of their VEVs are in principle achievable.
The same is true for the full constraints on the K\"ahler potential, see~\cite{Chen:2019ewa,Almumin:2021fbk},
including the computation of the potential of $T$ which corresponds to the ``evergreen'' problem of moduli stabilization, 
in the present context see~\cite{Novichkov:2022wvg} and references therein.
All these tasks have not been solved so far and also remain as open questions for our model. 

Finally, note that while our investigation here was focused on the flavor structure, the framework we work in 
has successfully been shown in many earlier influential works to be capable of a realistic phenomenology
also with respect to many other questions of particle physics and cosmology. Examples include
grand unification with symmetry based explanations for proton stability and the suppression of the $\mu$-term,
mechanisms for supersymmetry breakdown and a successful solution to the hierarchy problem, 
an origin of dark matter \textit{etc.}, see references in~\cite{Baur:2022hma}. It may be attractive to complete our construction also 
in the extension to other relevant phenomenological questions, such as identifying the cause of inflation, 
or the origin of the baryon asymmetry of the Universe.

\section{Summary}
There are explicit models of compactified heterotic string theory that reproduce in the IR the MSSM+eclectic flavor symmetry+flavon fields.
The complete eclectic flavor symmetry here can be unambiguously computed from the outer automorphisms of the Narain space group
and it non-trivially unifies previously discussed traditional, modular, R and \CP-type flavor symmetries, see equation~\eqref{eq:eclectic}.
The eclectic flavor symmetry is broken by vacuum expectation values of the moduli and of the flavon fields. 
While residual symmetries are common, their breaking and subsequent approximate nature can help to naturally generate
hierarchies in masses and mixing matrix elements. This allows analytic control over the generated hierarchies.

Here we have identified one example of a heterotic string theory model compactified on $\mathbbm{T}^2/\Z3$ that can give rise to a 
realistic flavor structure of quark and lepton sectors. To show this, we have derived the super- and K\"ahler potential and 
identified vacuua that give rise to non-linearly realized symmetries which allow to protect potentially realistic hierarchical flavor structures. 
Using the parameters of the effective superpotential, the non-canonical K\"ahler potential, as well as the vacuum expectation values of flavons
and the $T$ moduli field, we have performed a simultaneous fit to all experimentally determined quark and lepton sector parameters.
All observables can be accommodated and several to date undetermined parameters in the lepton sector are predicted by the fit. 
Nontheless, we stress that our goal was not primarily the derivation of these predictions (which are likely very model specific) 
but to demonstrate as a proof of principle that a realistic SM flavor structure can be 
obtained in the tightly symmetry constrained and predictive framework of UV complete string theory models.
Further topics to be investigated encompass the inclusion of the extra tori, the question of the computation of the flavon potential, as well as moduli stabilization.

\section*{Acknowledgements}\noindent
I would like to thank my collaborators Alexander Baur, Hans Peter Nilles, Saul Ramos-S\'anchez, and Patrick Vaudrevange.
A special thanks goes to Eleftheria Malami and Maria Laura Piscopo for supporting the experimental application of Monte Carlo methods 
during a visit to the Casino Baden-Baden.
\bibliography{Orbifold}
\bibliographystyle{JHEP}
\end{document}